\documentclass[prl,twocolumn,graphicx,amssymb,floatfix]{revtex4}

\usepackage{graphicx}
\begin{document}

\title{Reply to a Commentary “Asking photons
where they have been without telling them
what to say”}
\author{A. Danan, D. Farfurnik,  S. Bar-Ad, and L. Vaidman}
\affiliation{ Raymond and Beverly Sackler School of Physics and Astronomy\\
 Tel-Aviv University, Tel-Aviv 69978, Israel}

\begin{abstract}
Interesting objections to conclusions of our experiment with nested interferometers raised by Salih in a recent Commentary are analysed and refuted.
\end{abstract}

\maketitle
In a recent Commentary, Salih \cite{Sal} claims that we ``devised an elegant experiment investigating the past of
  photons inside two Mach-Zehnder interferometers, one inside the other - yet drew the wrong conclusions \cite{Danan}''. He also  argues that the story told  by the two-state vector formulation (TSVF) that we advocate, is  contradictory. Here we answer Salih's criticism.

Salih considers three possible options for the past of photons in our experiment and argues that option (1) according to which the photons are present in paths A and B simultaneously, is ruled out.   To support his claim he notices that the product of projections on A and on B vanishes.
 However, for pre- and post-selected systems, as the photons in our experiment, the product rule does not hold \cite{VHa}, and therefore, his argument fails.   The photon was in A and in B because it left traces in both places and this is the criterion of the past of the particle we rely on \cite{past}.  An unavoidable non-vanishing interaction with the environment leads to a ``weak measurement'' of the presence of the photon in various places inside the interferometer exhibiting  ``weak-measurement elements of reality'' \cite{WMER}.

\begin{figure}[t]
 \begin{center} \includegraphics[width=7.0cm]{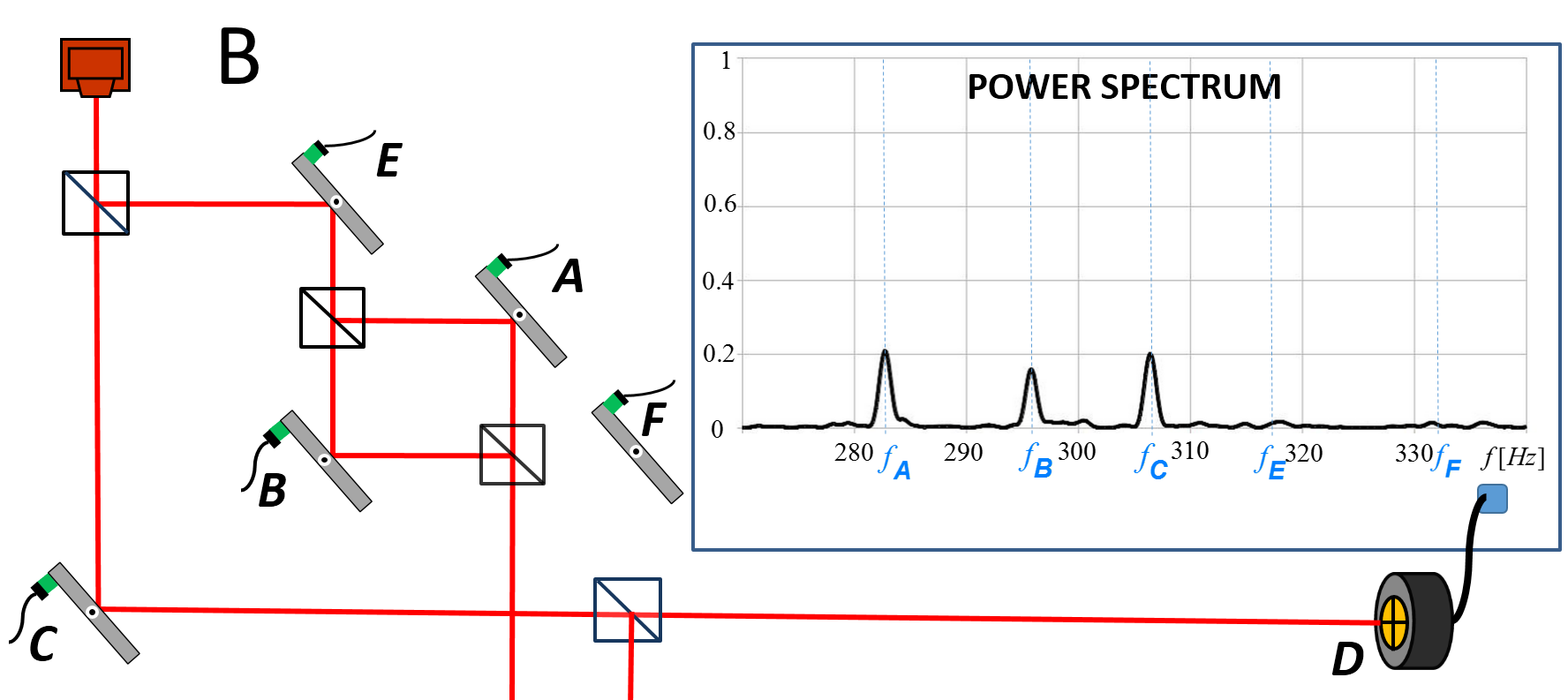}\\ \includegraphics [width=7.5cm]{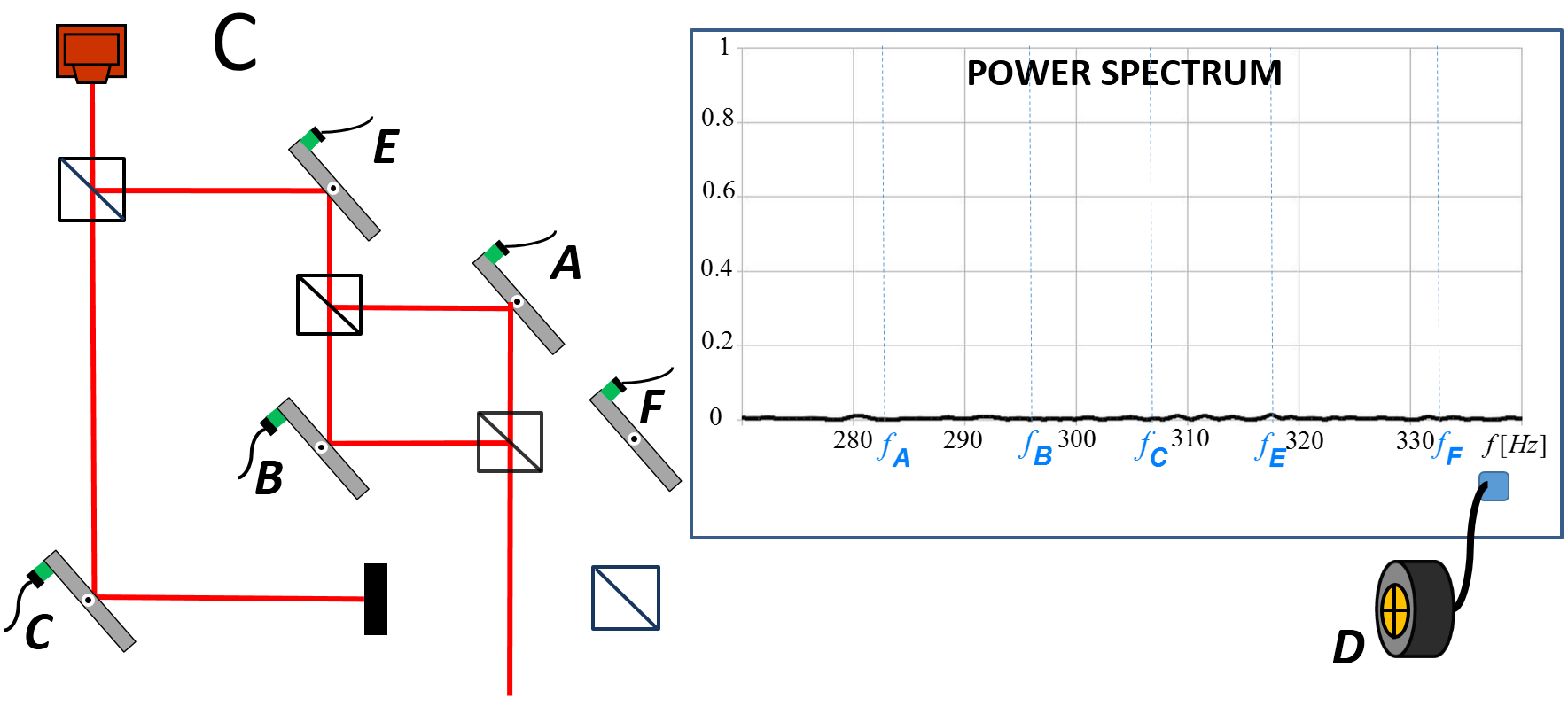}\\
      \caption{ (B) When the inner  interferometer is tuned in such a way that the beam of light passing through it does not reach the photo-detector, the power spectrum of the signal in the photo-detector still shows frequencies of the mirrors of this interferometer. (C) These frequencies (and all other signals) disappear when we block the lower arm of the large interferometer without changing anything in the upper arm. [This is Figure 2 of Ref. 2.]} \label{2}\end{center}
\end{figure}

Our claim, indeed, looks paradoxical. Even  if the photon left very small traces in both places,  there is a nonvanishing probability that the traces will be identified with certainty. In this case  a single photon will be found in two places simultaneously.  This is a contradiction: a single photon cannot be detected simultaneously in two places even in a non-demolition measurement. The resolution of the paradox is that the traces in A and in B are entangled and simultaneous detection of the photon in two places cannot happen. The photon changes the reduced density matrix in A  and, also,  the reduced density matrix in B, but, if this change is detected in A, then the reduced density matrix in B  becomes identical to the undisturbed density matrix there, and vice versa.

Salih considers a modification of our experiment in which, as he correctly states, we will not observe the presence of the photons in A and in B. However, it is not because the photons were not there, but because the modification spoils the experiment.  The weak value of the projection on A is 1. This means that  the effect on any measuring device weakly coupled to the photon in A will be  as if there was a single photon in A. The weak value of the projection on B is -1. It does not mean  that there is -1 photon, or that the photon has some ``negative probability'' for being in B. It means that the effect on any measuring device weakly coupled to the photon in B is as if there is one photon in B, but which has a special property of   coupling to everything with an opposite sign. The  presence of the photon in A shifts the position of the light on the quad-cell detector in one direction while the presence in B equally shifts it in the opposite direction, so the net shift is zero. Salih's modification transforms our experiment to weak measurement of the sum of the projections.
   The sum rule holds for weak values, so $({\rm \bf P}_A+{\rm \bf P}_B)_w=({\rm \bf P}_A)_w+({\rm \bf P}_B)_w=0 $. Introducing different frequencies of the mirror vibrations in our experiment led to  separation of measurements of the projection operators, so we were able to observe the presence of  photons in A, in B (and in C).

Salih correctly states that in any experiment in which the amplitude near mirror F is  exactly zero, as in his modification of our experiment, the presence of the photons near A and B will not be detected. This is obvious when we analyze the experiment using solely the forward evolving wave function.  As explained in our Letter \cite{Danan}, in this language,  it is the small leakage of the wave  through F which is responsible for the effect. But again, the fact that a particular experiment does not show the presence of the photon in A and B does not prove that the particle was not there.

Even if one performs local weak measurements of the presence of photons in A and in B with external measuring devices, the entanglement with these devices will spoil exact destructive interference and there will be some leakage of the amplitude towards F (and leakage of the amplitude of the backward evolving state towards E).
The traces in F and E vanish only in the limit of an ideal system without intermediate interactions or measurements. We state that the photons  were present in  A and B, but not in E and F because the ratio between the magnitudes of the traces goes to zero in the limit of  weak identical couplings in all places. Thus, in a weak measurement experiment only the traces in A, B (and C) will be observed,  as we have seen in our experiment.    Note, however, that the presence in E and F has   a different status relative to places outside the interferometer, so it might be helpful to define a ``secondary presence'' of the photon in these places \cite{morepast}.

Salih mentions that we have not provided the TSVF analysis of the experiment shown on our Fig. 2c (reproduced here),   the case when the channel $C$ was blocked.  Indeed, we brought this case  not as a  demonstration of the power of the TSVF, but for showing that our experimental results in Fig. 2b were not due to some  technical error of not properly screening   electronic signals.
 Yes, for Fig. 2c the simple TSVF analysis fails because  the weak values become  singular.  Since the post-selected state is orthogonal to the forward evolving state, formally, such post-selection is impossible, but the corresponding outcome of the final measurement might happen due
to imperfect optical elements and disturbances.   The TSVF analysis of this situation is less elegant, but possible,   see \cite{Sud,Pang}, and it does provide a way to calculate the size of the peaks which were too small to observe in our experiment.

In summary, Salih is correct that in his modification of the experiment the presence of  photons in the inner interferometer will not be detected. We argue, however, that the photons are there, Salih's experiment is just not a proper way to observe them. Any experiment which leaves traces in A and B (including ours) leaves some tiny traces in E and F. Salih's suggestion that ``the proper'' experiment must have exact zero amplitude in F does not correspond to a real  world in which   there is always some local interaction which leads to nonvanshing (albeit sometimes very small) amplitudes in all paths of the interferometer.

 The axioms of standard quantum mechanics do not tell us ``where is the particle''. Vaidman's proposal \cite{past} to define the past of a photon as places where it leaves a weak trace is consistent and it has been demonstrated in our experiment, including the surprising result of the presence of the trace inside the inner interferometer  without observable traces in the paths leading towards or away from it.

This work has been supported in part by the Israel Science Foundation  Grant No. 1311/14  and the German-Israeli Foundation  Grant No. I-1275-303.14.

\end{document}